\documentclass[aps,pra,twocolumn,superscriptaddress,preprintnumbers,
amsmath,amssymb,floatfix]{revtex4}

\usepackage{graphicx}
\usepackage{braket, amsfonts}
\usepackage{amssymb}
\usepackage{color}
\usepackage{hyperref}

\newcommand{\e}{\text{e}}

\begin{document}


\title[X-ray-frequency modulation via periodic switching of an external magnetic field]{X-ray-frequency modulation via periodic switching of an external magnetic field}


\author{Gregor Ramien}
\affiliation{Max-Planck-Institut f\"ur Kernphysik, Saupfercheckweg 1, D-69117 Heidelberg, Germany}

\author{Jonas Gunst}
\affiliation{Max-Planck-Institut f\"ur Kernphysik, Saupfercheckweg 1, D-69117 Heidelberg, Germany}

\author{Xiangjin Kong}
\email{xjkong@mpi-hd.mpg.de}
\affiliation{Max-Planck-Institut f\"ur Kernphysik, Saupfercheckweg 1, D-69117 Heidelberg, Germany}

\author{Adriana P\'{a}lffy}
\email{palffy@mpi-hd.mpg.de}
\affiliation{Max-Planck-Institut f\"ur Kernphysik, Saupfercheckweg 1, D-69117 Heidelberg, Germany}


\date{\today}


\begin{abstract}
Single x-ray photons can be resonantly scattered and stored with the help of suitable transitions in the atomic nucleus. Here, we investigate theoretically means of mechanical-free modulation
for the frequency spectra of such x-ray photons via periodic switching of an external magnetic field. We show that periodically switching on and off an external magnetic field generating hyperfine splitting of the nuclear transition leads to the generation of equidistant narrow sidebands of the resonantly scattered response. This frequency-comb-like structure depends on the magnitude and orientation of the applied magnetic field and on the switching period. An analytical approach for the characterization of the comblike frequency spectrum is presented. The feasibility of the external control on the frequency modulation of the x-ray response is discussed in view of possible applications in high-resolution spectroscopy or quantum technology.

\end{abstract}




\maketitle


\section{Introduction}
The recent advent and commissioning of coherent x-ray sources such as the X-ray Free Electron Laser \cite{LCLS-web,SACLA} brings new developments for the  field of x-ray physics and in particular for x-ray quantum optics \cite{xqo}. While not yet as advanced as its optical counterpart, the latter may enable coherent control of x-rays, with potential applications for the fields of metrology, material science, quantum information, biology and chemistry. The desirable properties of x-rays are deeper penetration, better focus, no longer limited by an inconvenient diffraction limit as for optical photons, correspondingly spatial resolution, robustness, and the large momentum transfer they may produce. With suitable control, x-rays could become desirable quantum information carriers.

A peculiar circumstance is that x-rays are resonant to either inner shell electron transitions in (highly) charged ions, or transitions in atomic nuclei.  Nuclear transitions in the x-ray regime can be used to manipulate single x-ray quanta \cite{Palffy2009,Liao:2012,kong2016stopping,palffy4,freqcomb,heeg2015tunable,PhysRevLett.111.073601,Heeg2017}. The key for such control is the use of M\"ossbauer transitions in  solid-state targets which enable collective effects to come into play in the nuclear excitation and decay processes \cite{Hannon1999}. In nuclear forward scattering (NFS), x-rays incoming on an $^{57}\mathrm{Fe}$ sample lead to a single delocalized excitation, also known as nuclear exciton \cite{Hannon1999} and coherent scattering of the resonant photons in the forward direction. The coherent decay of the exciton is faster than the spontaneous decay known from a single $^{57}\mathrm{Fe}$ nucleus, being the dominant decay channel on time scales shorter than the excited state lifetime. It has been shown experimentally that this decay channel can be controlled externally for instance by magnetic field switchings \cite{Shvydko1996}. Mechanical modulation of the x-ray photon wavepacket has also been achieved \cite{freqcomb,PhysRevA94.043849,Heeg2017,Vagizov2017}. In addition, thin-film planar x-ray cavities with  embedded $^{57}$Fe nuclear layers have proved to be particularly successful in demonstrating and exploiting collective effects in x-ray single photon superradiance, for instance by the observation of the collective Lamb shift \cite{rohlsberger2010collective}, electromagnetically induced transparency \cite{rohlsberger2012electromagnetically}, slow light in the  x-ray regime \cite{heeg2015tunable} or proofs of the collective strong coupling of x-ray photons and nuclei \cite{haber2016,haber2017}.

Frequency, phase, and amplitude modulation of x-rays based on M\"ossbauer effects has been the subject of several studies in the last decades.
Phase modulation of nuclear transitions has been realized in several experiments using M\"ossbauer sources, for instance, an x-ray echo of the 14.4 keV transition was produced by applying a mechanical phase modulator \cite{PhysRevLett.66.2037} and sidebands in the spectra of M\"ossbauer $^{57}\mathrm{Fe}$ nuclei were generated either by the thermo-mechanical oscillations induced by the radio-frequency current \cite{PhysRevLett.66.1934} or by spin waves of large amplitude transported from ferromagnetic sources \cite{PhysRevLett.62.2547}. More recently, the coherent control of the single-photon wavepackets shape to generate a gamma-ray comb for high-resolution spectroscopy \cite{freqcomb,PhysRevA94.043849} and the spectral narrowing of x-ray pulses for precision spectroscopy \cite{Heeg2017}  have been demonstrated  using mechanical motion. In addition to $^{57}\mathrm{Fe}$ nuclei, phase modulation of $^{67}\mathrm{Zn}$ M\"ossbauer nuclei was also demonstrated \cite{PhysRevLett.60.643}. We note that mechanical frequency modulation using $^{57}\mathrm{Fe}$ nuclei \cite{freqcomb,PhysRevA94.043849} was so far the only successful method to generate an x-ray frequency comb, despite achievements in the XUV regime \cite{XUV2005,XUV2012} and a recent proposal aiming at a broadband high-resolution solution \cite{Cavaletto2014}.

In this work, we theoretically investigate means of mechanical-free x-ray frequency modulation via switchings of an external magnetic field. The starting point is  the storage effect  induced by abrupt switching off the magnetic field that generates  hyperfine splitting of the nuclear transition \cite{Liao:2012}. We show that  periodically switching on and off the magnetic field leads to the generation of equidistant narrow sidebands of the resonantly scattered response. This frequency-comb-like structure depends on the magnitude and orientation of the applied magnetic field and on the period of  its modulation. The magnetic switching scheme thus provides additional external control means on the frequency modulation of the x-ray response, with possible applications in high-resolution spectroscopy, more precise tests of astrophysical models, quantum electrodynamics, or the variation of fundamental constants and in quantum technology.

The paper is structured as follows. An overview on NFS and the setup under investigation is presented in Section \ref{setup}. A brief description of the theoretical approach based on the Maxwell-Bloch equations is given in Section \ref{theo}. Our numerical results for several magnetic switching schemes together with a semi-analytical approach to explain the existence and properties of the predicted  comb-like frequency spectrum are presented in Section \ref{numres}. We conclude with a word on the experimental feasibility of the magnetic field modulation and a short discussion. Atomic units ($\hbar=1$) are used throughout the paper unless otherwise specified.

\section{Setup \label{setup}}

\begin{figure}[!t]
\centering
\includegraphics[width=0.85\linewidth]{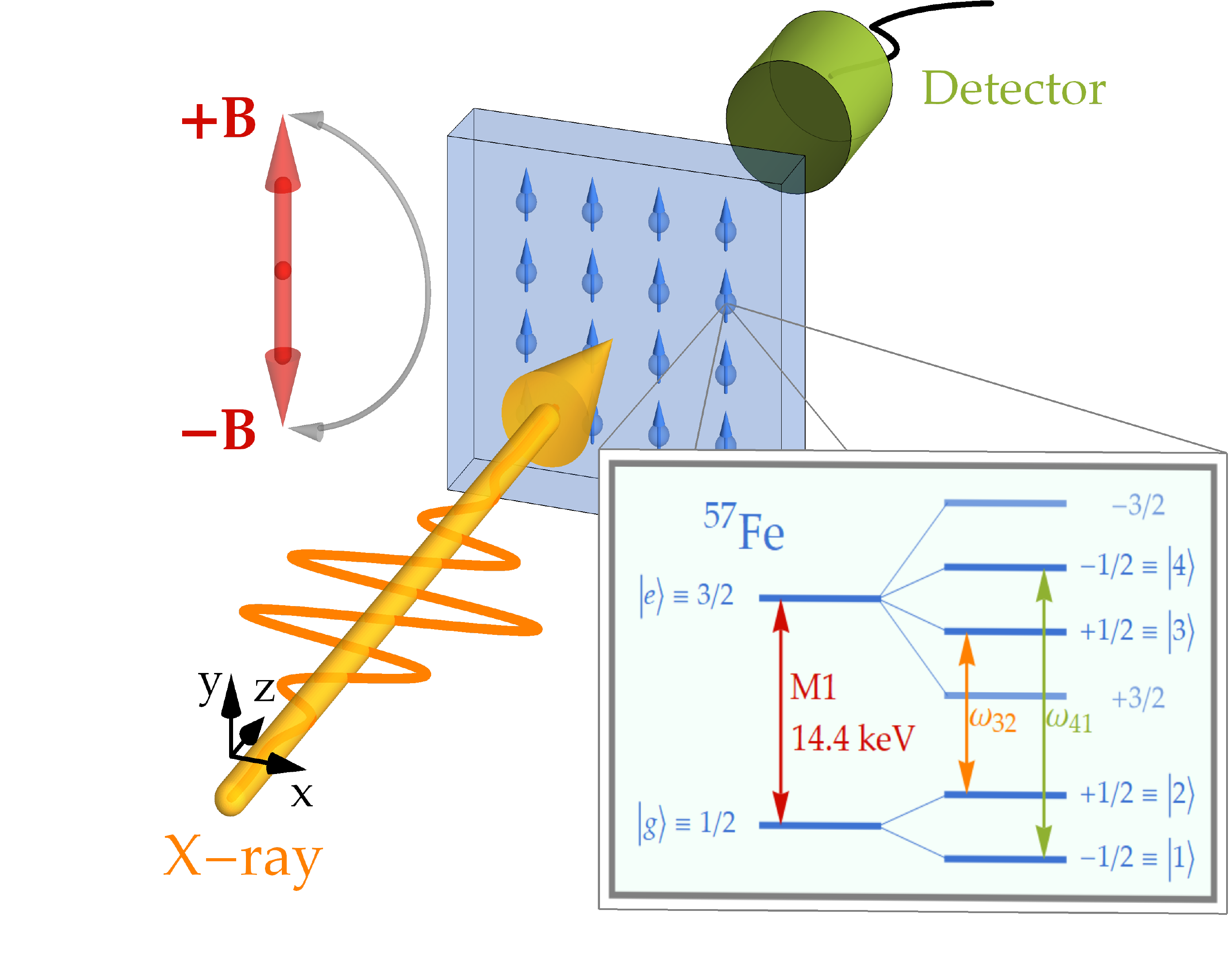}
\caption{NFS setup. The x-ray (orange line and arrow) propagating along the $z$-direction impinges on the M\"ossbauer target. The probe experiences a magnetic field $\mathbf{B}$, that can be switched off, retrieved and can be rotated by $\pi$, inverting  the quantization axis. The inset shows the level scheme of the ground $g$ and first excited $e$ state of the $^{57}\mathrm{Fe}$ sample including the magnetic sublevels. }
\label{fig:setup}
\end{figure}

We consider a standard NFS setup as depicted in Fig.~\ref{fig:setup} with  x-ray pulses propagating along the $z$-axis  and impinging on a M\"ossbauer target containing $^{57}\mathrm{Fe}$. The interval between incident x-ray pulses, typically from a synchrotron radiation (SR) source, is chosen such that it is larger than $1/\Gamma$, where $\Gamma$ denotes the spontaneous decay rate of a single nucleus. The pulses are tuned to the nuclear transition frequency of 14.413 keV between the stable ground and the first excited states of $^{57}\mathrm{Fe}$. For SR, the spectral width of the incident x-ray radiation (on the order of meV after monochromatization) is much larger than the nuclear transition width of 5~neV. Thus, each
pulse contains typically at most one photon resonant to the nuclear transition. The resonant photon is most likely absorbed by the nuclei in the sample, with an attenuation length of approx.~20~$\mu$m. The delayed nuclear response is then recorded in the forward direction, as illustrated schematically in Fig.~\ref{fig:setup}.

The 14.413 keV M\"ossbauer nuclear transition in $^{57}\mathrm{Fe}$ is a magnetic dipole ($M1$) transition connecting the  ground state, characterized by spin $I_g=1/2$, and the first excited state with  $I_e=3/2$. Under the effect of an external magnetic field, the ground and the excited state split according to the magnetic quantum numbers $m_g = -1/2,1/2$ and  $m_e = -3/2,-1/2,1/2,3/2$ as shown in the inset of Fig.~\ref{fig:setup}.  The choice of polarization and magnetic field orientation determines which of the six possible $M1$ transitions between the two ground and four excited magnetic sublevels are driven \cite{Palffy2010}. For instance, choosing
 the incident beam to be linearly polarized along the $x$-axis while the magnetic field is parallel to the $y$-axis imposes the selection rule  $\Delta m=0$. Consequently, only two hyperfine transitions may occur and the system is effectively reduced to four levels: $m_g = \pm 1/2$ and $m_e = \pm 1/2$.
%

The resonant fraction of the x-ray pulses, i.e., the single photon, can be stored by manipulations of the hyperfine magnetic field. Ref.~\cite{Shvydko1996} demonstrated experimentally the storage and subsequent release of the x-ray photons by rotations of the hyperfine magnetic field. Such  abrupt rotations could be realized on a timescale faster than 4~ns \cite{Shvydko1994.EPL} in $^{57}$Fe-enriched FeBO${}_3$ \cite{vanBurck1987.PRL,Smirnov1983.JETP} due to the special magnetization properties of the latter. Refs.~\cite{Liao:2012,kong2016stopping} proposed a more efficient coherent storage by switching off the magnetic field. Additionally,  also phase modulation by inverting the orientation of the $\mathbf{B}$ field when switching the magnetic field back on was proposed. Here we investigate the effect of successive periodical switchings of the hyperfine magnetic field and show that they give rise to equidistant sidebands in the shape of an x-ray frequency comb. The theoretical tools to calculate the scattered spectra for varying magnetic fields are given by the Maxwell-Schr\"odinger or the Maxwell-Bloch equations (MBE) \cite{Scully2006} that are briefly introduced in the next section.


\section{Theoretical approach \label{theo}}
In this work we adopt a  general approach from atomic
quantum optics based on the MBE to determine the field propagation through the nuclear medium  taking into
account additional perturbations such as time-dependent magnetic fields. Since this approach adapted for NFS setups has been the subject of other works \cite{Kong:2014}, we only briefly sketch here the main equations that are used to obtain our numerical results.

We treat the interaction of the x-ray field with the nuclear sample semi-classically, i.e., considering the nucleus as a quantum system and the electromagnetic field classically.  The semi-classical approach is endorsed by experimental agreement \cite{shvydko1998, buerck1999}. The Bloch equations part from the MBE are deduced from the master equation \cite{Scully2006}
\begin{equation}
\partial_t\hat{\rho}=-i[\hat{H},\hat{\rho}] + \hat{\rho}_s\, ,
\label{eq:Master}
\end{equation}
with $\hat{H}$  the total Hamiltonian of the system consisting of the quantum system and the incident electromagnetic field $\textbf{E}(z,t)$, $\hat{\rho}$ the density matrix and the decoherence matrix $\hat{\rho}_s$ which is introduced to include quantum effects such as decoherence and spontaneous decay. The interaction part of the Hamiltonian is given in the radiation gauge by \cite{palffy:2008}
\begin{equation}
\hat{H}_1=-\frac{1}{c} \hat{\textbf j}(\textbf{r},t)\textbf{A}(\textbf{r},t),
\label{eq:H1}
\end{equation}
with  $\hat{\textbf j}$ being the nuclear current operator  and  $\textbf{A}$  the vector potential of the incident field, respectively, while $c$ denotes the speed of light. By identifying the multipole operator in the interaction Hamiltonian \cite{ringschuck:1980}, the matrix element for the latter can be written as \cite{Kong:2014,palffy:2008}
\begin{equation}
\label{eq:FinMatEl}
\begin{split}
\Braket{e|\hat{H}_1|g} \sim E_k(z,t) \;  (-1)^{I_g-m_g}C(I_e I_g L; m_e -m_g M) \\
\times \sqrt{\frac{2\pi(L+1)}{L}}\frac{k^{L-1}}{(2L+1)!!} \; \sqrt{\mathcal{B}(\mathcal{E}/\mathcal{M}L,I_g \to I_e)}\; ,
\end{split}
\end{equation}
where we consider a unidirectional propagation of the electric field along the $z$-axis $E_k(\textbf{r},t)=E_k(z,t)$. The notation $C(I_e I_g L; m_e -m_g M)$ used in the expression above stands for the Clebsch-Gordan coefficient \cite{edmonds:1994} coupling angular momenta $I_e$ and $I_g$, and $L$ is the multipolarity of the transition. Furthermore, $k=\omega_k/c$ is the wave vector of the electric field and $\mathcal{B}(\mathcal{E}/\mathcal{M}L,I_g \to I_e)$ is the nuclear reduced transition probability \cite{ringschuck:1980} for the nuclear transition of multipolarity $L$ and type $\mathcal{E}$ (electric) or $\mathcal{M}$ (magnetic). The Rabi frequency $\Omega$ is  defined here starting from the matrix element above as $E_k(z,t)\sqrt{\frac{2\pi(L+1)}{L}}\frac{k^{L-1}}{(2L+1)!!} \; \sqrt{\mathcal{B}(\mathcal{E}L,I_g \to I_e)}=: \Omega(z,t)$.

To include the sample's response, we need to consider the wave equation of the incident field determined by the Maxwell equations. Assuming the incident field $\textbf{E}$ as well as the source term current $\textbf{J}$ to be unidirectionally propagating in $z$-direction $\textbf{J}(z,t)=J(z,t)e^{i(k\cdot z - \omega_k t)} \textbf{e}_x$, we additionally apply the slowly-varying envelope approximation considering $|\partial_z^2 E| \ll |k \partial_z E|$, $|\partial_t^2 E| \ll |\omega_k \partial_t E|$ and $|\partial_t J| \ll |\omega_k J|$. The current density for a single nuclear resonance may be obtained by summing over all nuclei involved in the coherent scattering and tracing over the product $\mathbf{\hat{j}}(\textbf{k})e^{-ikz}\hat{\rho}$ of the current density operator (in momentum representation) and the density matrix \cite{Kong:2014,shvydko:1999}. This product can be connected to the interaction Hamiltonian in its alternative form $\hat{H}_1=\frac{i}{\omega_k} \mathbf{\hat{j}}^*(\mathbf{k}) E_k(z,t)\mathbf{e}_x e^{i(kz-\omega_k t)}$ \cite{shvydko:1993} such that we obtain the following expression for the  wave equation \cite{Kong:2014}:
\begin{equation}
\label{eq:WEfinal}
\left( \partial_z+\frac{1}{c}\partial_t \right)\Omega(z,t) =
 i \eta \big[ C_{32}\cdot\rho_{32}(z,t) - C_{41}\cdot\rho_{41}(z,t) \big].
\end{equation}
The parameter $\eta$ is given by $\eta:=\xi \Gamma/L$ with $L$ the length and $\xi$ the optical thickness of the sample, respectively. Furthermore, $C_{32}$ and $C_{41}$ are the corresponding Clebsch-Gordan coefficients for the two considered transitions, while the $(-1)^{I_g-m_g}$ factor in Eq.~(\ref{eq:FinMatEl}) has been written explicitely. We note that this notation differs from the one  used in Refs.~\cite{Liao:2012,Kong:2014}, where the definition of the coefficients included the sign factor.

The wave equation (\ref{eq:WEfinal}) together with the Bloch equations resulting from Eq.~(\ref{eq:Master}) are solved consistently for $\Omega(z,t)$ to evaluate the time-dependent NFS scattered pulse intensity $I\sim |\Omega(z,t)|^2$. The corresponding NFS frequency spectra are obtained  by a discrete Fourier transform of the forward scattered field evaluated at $z=L$.

%
%
\section{Numerical results and discussion \label{numres}}
Applying the formalism outlined above we may proceed to investigate the effects of periodic magnetic switchings on the NFS spectrum. Within certain constraints on the magnetic switching times, the field modulations lead to periodic sequences of photon storage and retrieval events. By analyzing the NFS spectrum in the time as well as frequency domain, we find that the periodicity in the $\mathbf{B}$-field switching generates comb-like frequency spectra with equidistant peaks in the forward scattered radiation. The comb-like shape can be controlled via the periodic properties of the magnetic switching.

\begin{figure}
\centering
\includegraphics[width=0.85\linewidth]{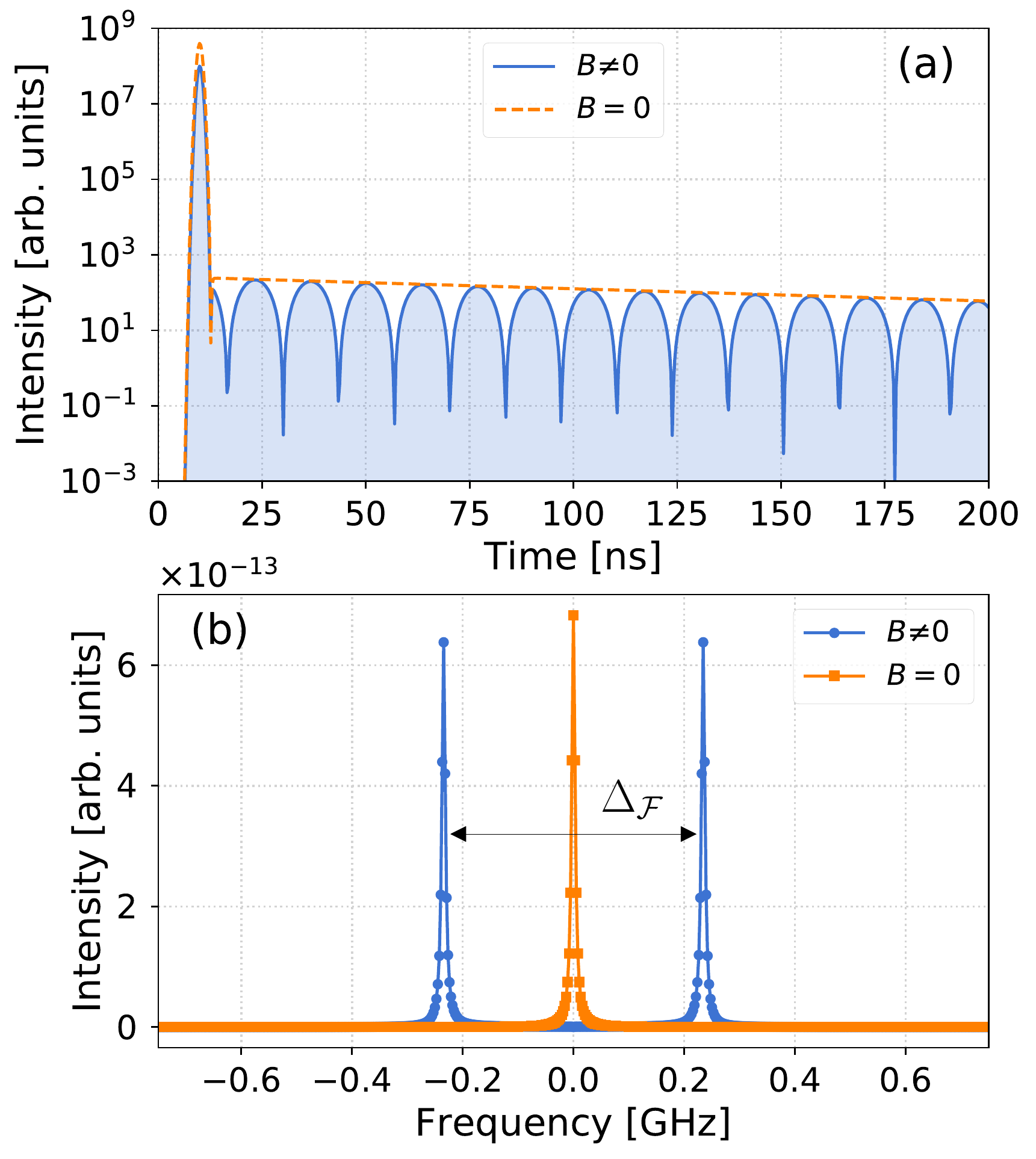}
\caption{(a) Time spectrum for a constant magnetic field and optical thickness $\xi=0.5$ (blue solid line) and in the absence of any hyperfine magnetic field with $\xi=0.25$ (orange dashed line). The peak at $t=10$~ns shows the incoming Gaussian pulse. The  magnetic field introduces an absolute-cosine-like oscillation known as quantum beat. (b) Corresponding frequency spectrum. In the absence of the magnetic field, the spectrum consists of a single peak at the nuclear transition frequency $\omega_0$. The magnetic field introduces the splitting into the
 two transition frequencies $\omega_{41}$ and $\omega_{32}$ separated by $\Delta_\mathcal{F}$.}
\label{fig:NFS_spectra}
\end{figure}

We begin our discussion by analyzing the more simple case of NFS in the presence of constant magnetic fields. In  Fig.~\ref{fig:NFS_spectra}(a) we illustrate our numerical results for the temporal  forward scattering response for a constant magnetic field with $B_0=34.4$ T and $\xi=0.5$. The value of the magnetic field strength is chosen according to the intrinsic magnetic field at the nuclear site in FeBO${}_3$ crystals.
For comparison we also present the case without magnetic field ($B=0$) where the optical thickness has been adapted to $\xi=0.25$ to compensate for the different initial condition for the density matrix  \cite{Kong:2014}.
The time spectra are obtained by numerically solving the MBE given in Eqs.~(\ref{eq:Master}) and (\ref{eq:WEfinal})  considering a Gaussian-like incident pulse with a time delay of $t_{{d}}=10$
 ns and a pulse duration of $\tau=1$ ns: $\Omega(0,t) \propto \exp[-(t-t_{{d}})^2 / \tau^2]$.
The nuclei are assumed to be initially in the ground state with equal population among the hyperfine  sublevels: $\rho_{ij}(z,0) = 0.5 \delta_{ij}$ for $i,j \in \{1,2\}$ and $\rho_{ij}(z,0) = 0$ for $i,j \in \{3,4\}$.

In the absence of any external magnetic field, all magnetic sublevels are degenerate. The resonant driving of the nuclear transition leads to a decay which for the chosen small optical thickness $\xi=0.5$ closely resembles the simple natural, spontaneous decay  $|\Omega|^2 \propto \e^{-\Gamma t}$, as  shown in Fig.~\ref{fig:NFS_spectra}(a). In contrast, in the presence of a magnetic field the degeneracy of the energy levels is lifted, so that the ground state splits into two and the excited into four distinctive sublevels. With the geometry discussed in Section \ref{setup} and illustrated in Fig.~\ref{fig:setup}, the two $\Delta m =0$ transitions with energies $\omega_{32}$ and $\omega_{41}$ will be driven by the broadband SR pulse leading to a beat pattern in the temporal spectrum.
We note that the interference between the two transitions can only occur due to the delocalized nature of the excitation, as discussed in Ref.~\cite{eraser}. The uneven depth of the interference minima on the logarithmic scale is related to the limited numerical accuracy of the NDSolve Mathematica routine  used to numerically  solve the MBE.
The angular beat frequency $\omega_{QB}$  of the so-called quantum beats shown in Fig.~\ref{fig:NFS_spectra}(a) is given by
\begin{equation}
\omega_{QB} = \frac{\omega_{41}-\omega_{32}}{2} =  \frac{1}{2} \Delta_{B_0}.
\end{equation}
We introduce the notation  $T_{QB}$ for the quantum beat period, which corresponds to two beats, such that between two minima we have   $T_{QB}/2=2 \pi/\Delta_{B_0}$. As the hyperfine splitting is proportional to the magnetic field intensity, we can  narrow the quantum beat pattern by increasing the value $B_0$. We consider only cases with $\Gamma<\Delta_{B_0}$ to prevent too complex beat structures.

To obtain the corresponding NFS frequency spectra, we transform the numerical solution of the MBE  to the  frequency domain by a discrete Fourier transform of the forward scattered field evaluated at $z=L$: $\Omega(L, t) \mapsto \tilde{\Omega}(L, \omega) := \mathcal{F}[\Omega(L, t)]$. The length of the time range considered for the transform determines its resolution, the larger the range the higher the resolution in the frequency spectrum. To remove the contribution of the incident pulse itself and also of prompt scattering processes other than the resonant NFS, we only transform the interval starting from the first quantum beat minimum up to the maximum range of the temporal spectrum delivered by the numerical solution of the MBE \cite{Scully2006}. In the following we employ a time window of 2410 ns for the numerical solution of the MBE.  Together with the magnetic field strength of 34.4 T this leads to a discretization step of $\Delta f = 2.63$ MHz. In practice, the time window corresponds to the measurement time over which the periodic magnetic switching takes place. According to Heisenberg's uncertainty relation for time and energy, resolving the narrow peaks in energy can only be achieved by observation over a correspondingly long time interval.

The frequency spectra with and without hyperfine splitting are presented in  Fig.~\ref{fig:NFS_spectra}(b).
The frequencies are relative to $\omega_k = \omega_0$ (at zero detuning $\Delta=0$) due to a transformation into the rotating frame in the MBE \cite{Scully2006}.
As expected, for $B=0$ only the principal transition frequency $\omega_0$ occurs. With a constant magnetic field $B \neq 0$, we observe the frequencies of the two hyperfine-split transitions $\omega_{41}$ and $\omega_{32}$.
Since the hyperfine splitting $\Delta_{B_0}$ equals twice the frequency $\omega_{QB}$, we expect the peak separation $\Delta_\mathcal{F}=2\omega_{QB}$ to be given by $\Delta_{B_0}$ which evaluates to 0.4691 GHz in the case of $B=34.4$ T.
Numerically we obtain $\Delta_\mathcal{F} \approx 0.4689$ GHz which deviates from $\Delta_{B_0}$ by 0.02\%. Considering the limited frequency resolution due to the  recording window in the time domain, $\Delta_\mathcal{F}$ and $\Delta_{B_0}$ coincide within the expected error range of 0.0037 GHz.

As was shown in Ref.~\cite{Liao:2012}, switching off the hyperfine magnetic field at any minimum of the quantum beat should lead to coherent storage of the x-ray photon, strongly suppressing the scattered signal. A subsequent restoring of the magnetic field after an arbitrary time interval (within the lifetime of the excited state) will lead to the revival of the NFS signal. We will make use in the following of these two features to implement magnetic field modulations.

\subsection{Magnetic switching \label{B-field}}
We now proceed to analyze the resonantly scattered field in the presence of  periodically modulated magnetic fields. The magnetic field is abruptly switched off and on, considering two scenarios:

\begin{itemize}
\item  regular switching, alternating between the magnetic field at full strength  $B_0$ and zero,
\item inverted switching, alternating between $B_0$, $0$ and $-B_0$, i.e, switching the magnetic field back on  oriented in the opposite direction.
\end{itemize}
The critical parameters for these sequences are the time intervals of the different stages. We denote the storage time during which the magnetic field is zero as $\tau_{\rm{off}}$. As discussed in Ref.~\cite{Liao:2012}, this period can be chosen without any constraints.
The interval over which the $\mathbf{B}$ field is on is denoted by $\tau_{\rm{on}}$. If we aim at achieving maximum signal suppression by switching off the magnetic field, $\tau_{\rm{on}}$ should be restricted to multiples of $T_{QB}/2$, corresponding to switching off the magnetic field at the quantum beat minima.

The two stepwise switching schemes $B(t) = B_0 f(t)$ are implemented as sums of the rectangle function $\Pi$ which is defined in terms of the Heaviside Theta function $\Theta$ for a dimensionless variable $x$ as $\Pi(x) = \Theta(x+\frac{1}{2}) \Theta(\frac{1}{2}-x)$.
In the regular switching case, the sequence $f(t)$ is given by
\begin{equation}
f(t) = \Theta(t_0 - t) + \sum_{i=1}^{n} \Pi \left( \frac{t - t_0 - i (\tau_{\rm{on}} + \tau_{\rm{off}}) + \tau_{\rm{on}}/2}{\tau_{\rm{on}}} \right),
\label{eq:f_regular}
\end{equation}
where $\Theta(t_0 - t)$ ensures that the magnetic field is turned on initially and is switched off for the first time in the the first quantum beat minimum $t_0$. The summation limit $n$ is determined by the length of the simulated spectrum. For the inverting switching case we have
\begin{equation}
\label{eq:f_inv}
\begin{split}
f(t) = \Theta(t_0 - t) + \sum_{i=1}^{n} \Pi \left( \frac{t - t_0 - i (\tau_{\rm{on}} + \tau_{\rm{off}}) + \tau_{\rm{on}}/2}{\tau_{\rm{on}}} \right) \\
\times \left[ 1 - 2\, mod_2(i) \right] ,
\end{split}
\end{equation}
with the modulo map $mod_2$ for divisor $2$.

Note that throughout this section we set the optical thickness $\xi$ sufficiently low to suppress multiple scattering processes. This leads to a quantum beat envelope approximately given by the exponential decay $e^{-\Gamma t}$ \cite{shvydko:1999} as shown in Fig.~\ref{fig:NFS_spectra} and the beat lopes become equidistant, so that periodic switching sequences aligned with the moments of the quantum beat minima are easier to implement.


\subsection{Regular stepwise switching scheme}

\begin{figure}
\centering
\includegraphics[width=0.85\linewidth]{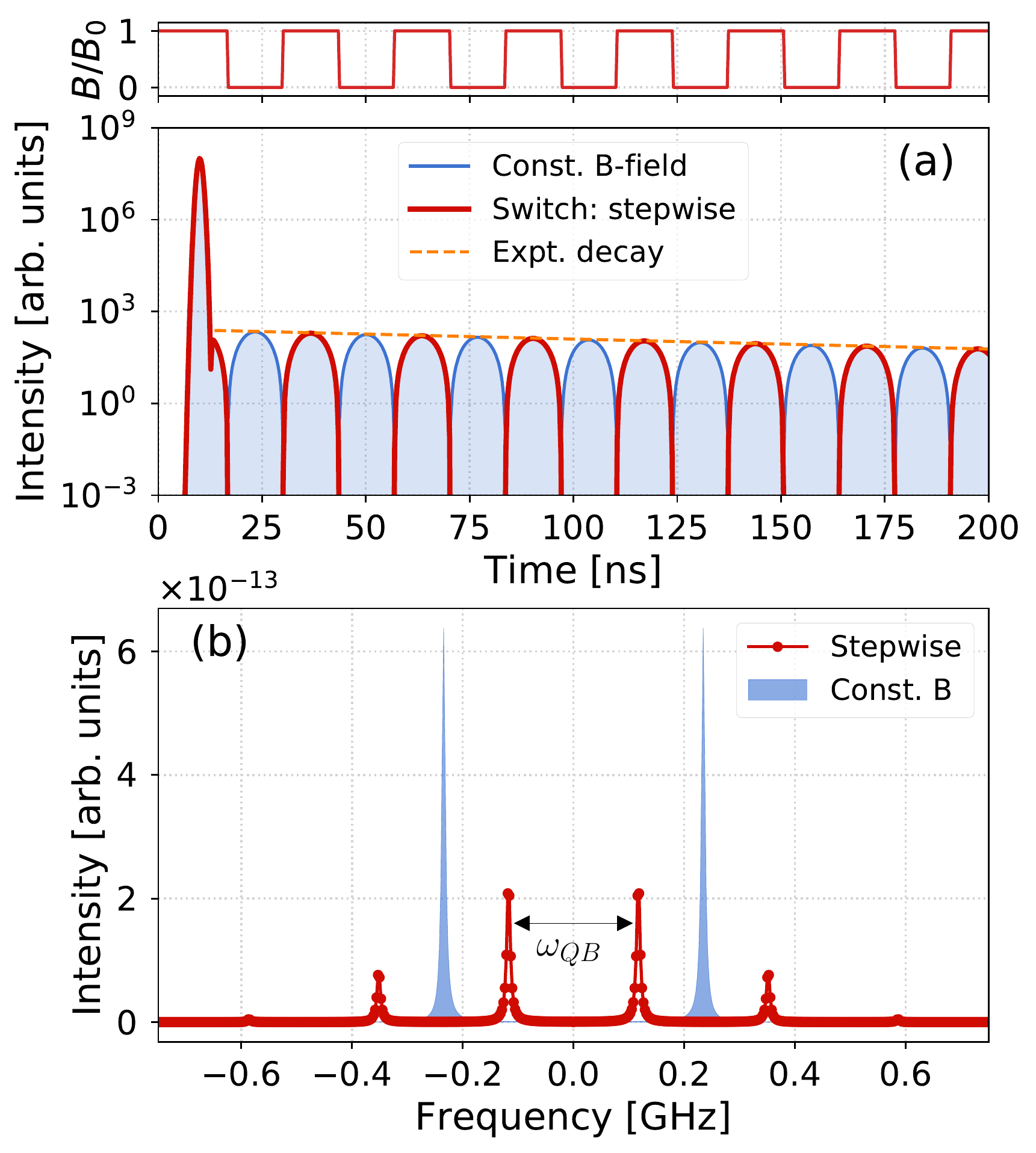}
\caption{(a) Time  and (b) frequency  spectra for $\tau_{\rm{on}}=\tau_{\rm{off}}=\frac{T_{QB}}{2}$ (thick red line). The periodically manipulated scattered field intensity is presented together with the case of a constant magnetic field $B_0=34.4$ T (blue line) and of an approx. exponential decay in the absence of the magnetic field (dashed orange line). For the computations an optical thickness of $\xi=0.5$ was used.}
\label{fig:stepwise}
\end{figure}

We solve the MBE numerically for a regular stepwise magnetic field switching sequence and different storage times in combination with a fixed $\tau_{\rm{on}}$ of half a QB period. The value $\xi$ is set to $0.5$ and $B_0$ to $34.4$ T. The numerical results for  $\tau_{\rm{off}}=T_{QB}/2$  are shown in Fig.~\ref{fig:stepwise} together with the case of constant magnetic field $B_0 = 34.4$ T. Fig.~\ref{fig:stepwise}(a) illustrates the pulse train (with exponentially decaying envelope) generated by the periodic magnetic field switching choosing $\tau_{\rm{on}}=\tau_{\rm{off}}=T_{QB}/2$ and switching off the magnetic field always at the minima of the quantum beat. The interval between each pulse evaluates to $\tau_{\rm{on}} + \tau_{\rm{off}} = T_{QB}$. The scattered signal during the periods with no magnetic field present is stronly suppressed. The steep switching is numerically challenging but fortunately numerical artifacts only affect the spectrum regions during $\tau_{\rm{off}}$  which are about six orders of magnitude lower than the corresponding $\tau_{\rm{on}}$ intervals.

The corresponding frequency spectrum shown in  Fig.~\ref{fig:stepwise}(b) reveals that under the influence of the periodic $\mathbf{B}$-field switching, the two peaks with spacing $\Delta_{\mathcal{F}}=2\omega_{QB}$ shown in Fig.~\ref{fig:NFS_spectra}(b) split into several spikes with spacing $\omega_{QB}$. Thus, the periodicity of the magnetic switching is transferred to the resonantly scattered radiation resulting in a comb-like pattern in the frequency domain.

\begin{figure}
\centering
\includegraphics[width=1.0\linewidth]{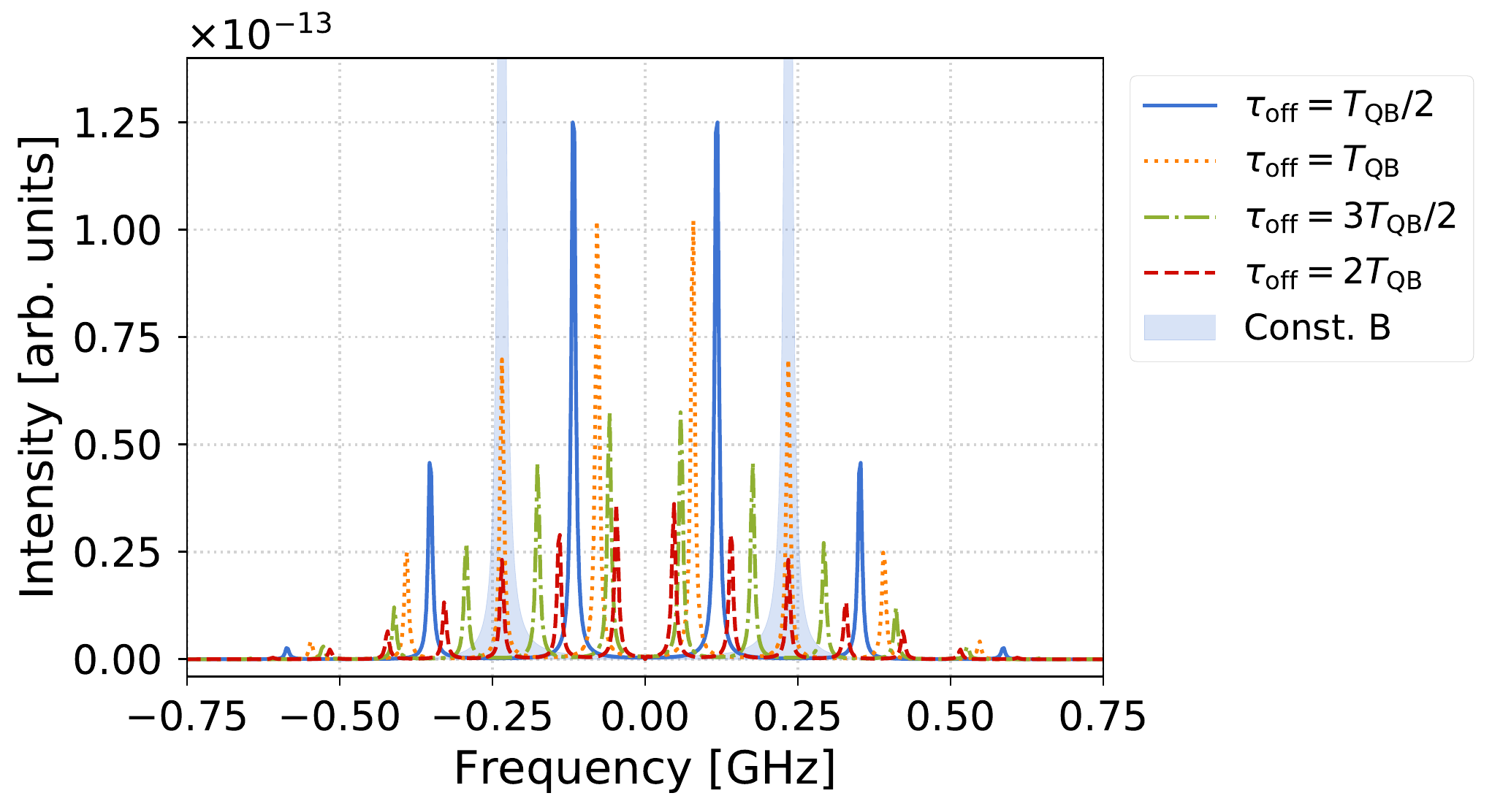}
\caption{Frequency spectra for different $\tau_{\rm{off}}$ values and constant $\tau_{\rm{on}}=T_{QB}/2$. With increasing storage time, the peaks multiply and narrow towards $\omega_0$ while the overall intensity drops. For illustration purposes, the spectrum for $\tau_{\rm{off}}=T_{QB}/2$ was manually downscaled by 40\%.}
\label{fig:tau_off_stepwise}
\end{figure}

In order to study this behavior more quantitatively we have performed calculations for varying storage times $\tau_{\rm{off}}$ with resulting frequency spectra presented in Fig.~\ref{fig:tau_off_stepwise}.
From this analysis we can conclude that at constant $\tau_{\rm{on}}=T_{QB}/2$ the separation of the peaks in the frequency combs narrows with growing $\tau_{\rm{off}}$. The peaks drift towards the origin ($\omega_0$) and new peaks gain intensity at the border regions (towards $\pm 0.75$ GHz). The two frequencies $\omega_{41}$ and $\omega_{32}$ appear in the combs for every $\tau_{\rm{off}}$ set to an even multiple of $\frac{T_{QB}}{2}$. Due to maximum NFS signal suppression during $\tau_{\rm{off}}$, the integrated intensity decreases as we increase this parameter. We have also performed further calculations for constant $\tau_{\rm{off}}$ but varying $\tau_{\rm{on}}$ (while still switching off the magnetic field always at the minima of the quantum beat) that show a similar appearance of comb-like structures in the frequency domain.  As it will be discussed in detail in Section~\ref{analysis}, the peak separation depends on both $\tau_{\rm{off}}$ and $\tau_{\rm{on}}$, or  more precisely on the quantity $(\tau_{\rm{on}}+\tau_{\rm{off}})$. Further calculations show that the comb-like structure persists also for longer Gaussian input pulses, i.e., for pulse duration $\tau=10$ and 20~ns, considering $t=\tau$ as the start of the $\mathbf{B}$-field modulation. For input pulses provided by the exponential decay of  a M\"ossbauer source, magnetic field switchings implemented during the decay also lead to the appearance of narrow peaks  in the frequency spectrum of the scattered x-ray photon.

\subsection{Inverted stepwise switching scheme}
\begin{figure}
\centering
\includegraphics[width=0.85\linewidth]{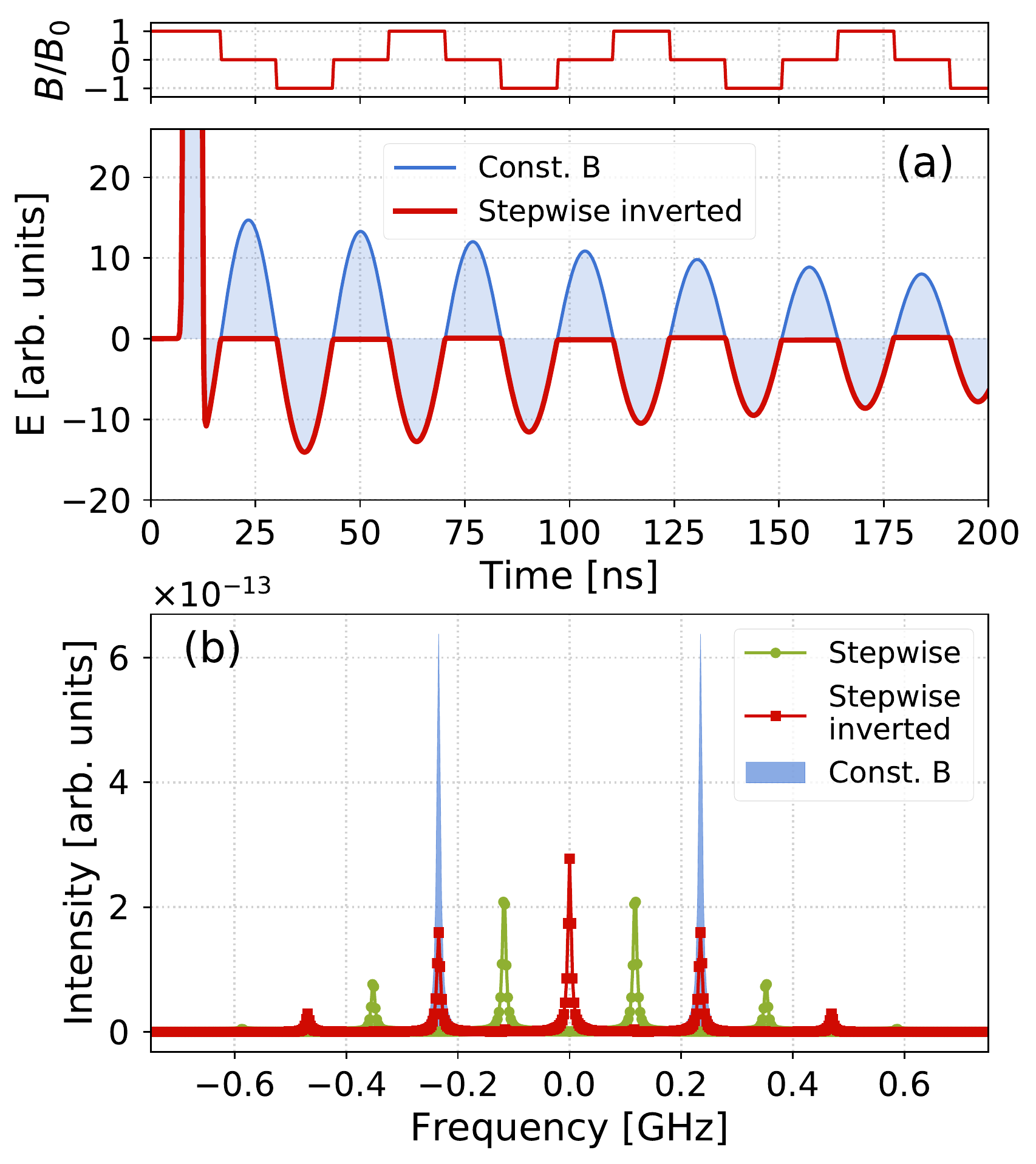}
\caption{(a) $\mathbf{E}$-field amplitude time spectrum for the inverting sequence at $\tau_{\rm{on}}=\tau_{\rm{off}}=T_{QB}/2$ (thick red line). The magnetic field orientation is inverted after every storage. (b) Comparison of the frequency spectra of the inverting and regular stepwise switching for the same choice of  $\tau_{\rm{on}}$ and $\tau_{\rm{off}}$.}
\label{fig:stepwise_inverted}
\end{figure}

We investigate the effect of the magnetic field orientation by solving the MBE for the inverted stepwise switching schemes described in Section~\ref{B-field}.
This sequence differs from the regular one in that the orientation of the $\mathbf{B}$ field is inverted every time the field is switched on. This $180^\circ$ rotation causes a $\pi$-phase shift in the time domain as can be seen from the resulting $\mathbf{E}$-field amplitude presented in  Fig.~\ref{fig:stepwise_inverted}(a). The corresponding intensity spectrum is the same as the one presented in  Fig.~\ref{fig:stepwise}(a).

Interestingly, the Fourier spectrum for the inverted switching scheme differs significantly from the regular case as can be seen in the comparison presented in Fig.~\ref{fig:stepwise_inverted}(b) for $\tau_{\rm{on}}=\tau_{\rm{off}}=T_{QB}/2$. The inverted sequence also generates a comb-like structure in the frequency domain, however, with shifted  peaks relative to the regular outcome by half their equidistant separation. The highest peak appears at $\omega_0$. The distance between neighboring peaks is again given by $\omega_{QB}$.

\begin{figure}
\centering
\includegraphics[width=1.0\linewidth]{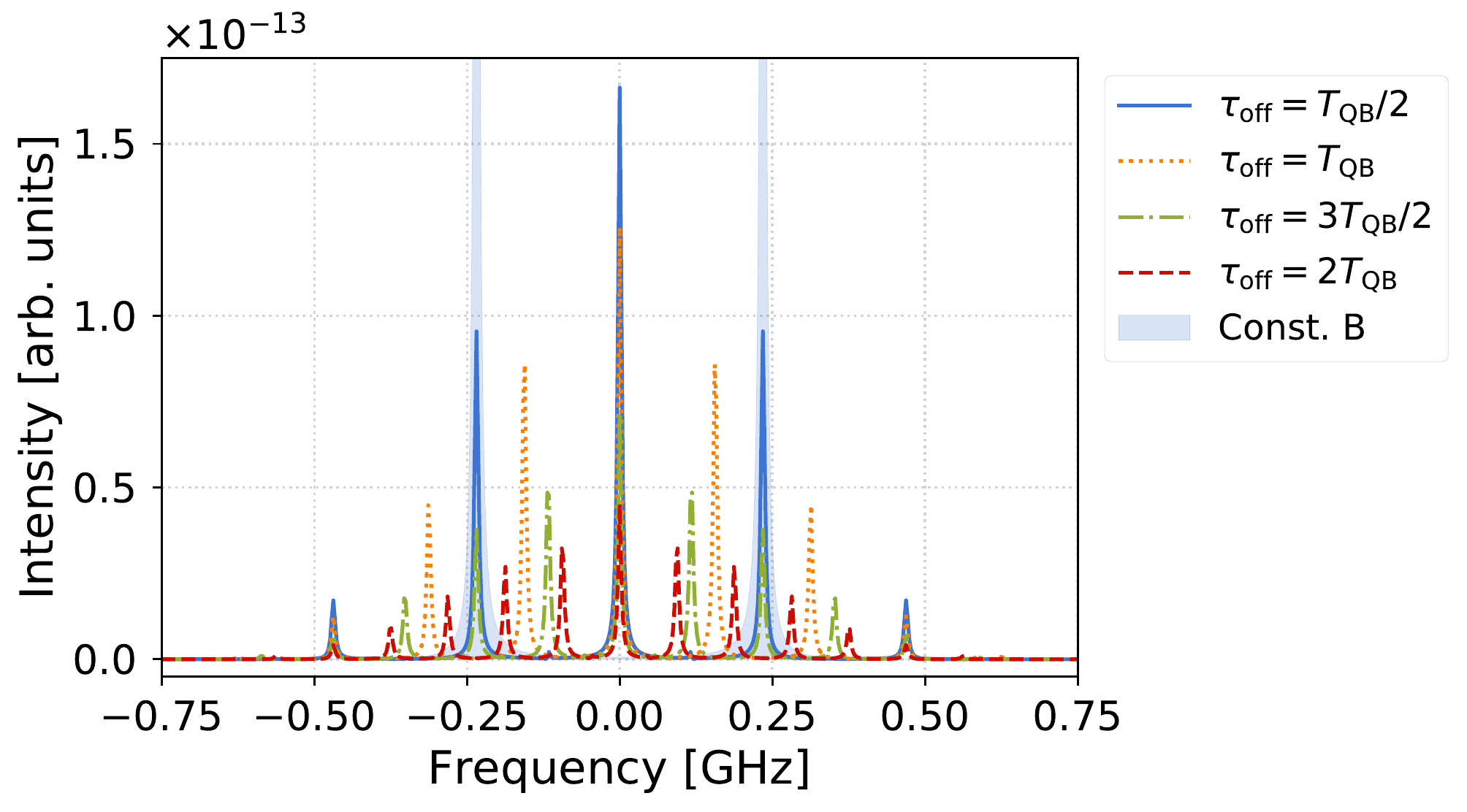}
\caption{Frequency spectra in the case of the inverted stepwise switching scheme for several storage times $\tau_{\rm{off}}$ and constant $\tau_{\rm{on}}=T_{QB}/2$. For illustration purposes, the spectrum for $\tau_{\rm{off}}=T_{QB}/2$ was manually downscaled by 40\%. }
\label{fig:tau_off_stepwise_inverted}
\end{figure}

In Fig.~\ref{fig:tau_off_stepwise_inverted} we present the implementation of the inverted sequence for several storage times $\tau_{\rm{off}}$ ranging from $T_{QB}/2$ to $2 T_{QB}$.
For varying $\tau_{\rm{off}}$ and fixed $\tau_{\rm{on}}$, the frequency spectra show a similar behaviour as for the regular switching. While the $\omega_0$-line is always present, peaks at $\omega_{41}$ and $\omega_{32}$ only appear if $\tau_{\rm{off}}$ is an odd multiple of half the quantum beat period $T_{QB}$. Moreover, for increasing $\tau_{\rm{off}}$ the peak separation $\Delta_{\mathcal{F}}$ reduces and new higher frequency components appear.

Another remarkable effect comes about for vanishing storage time $\tau_{\rm{off}}\rightarrow 0$. This corresponds to periodic rotations of the $\mathbf{B}$-field orientation with period $\tau_{\rm{on}}$.
The phase shift results in an $\mathbf{E}$-field time spectrum that resembles an absolute cosine. Since the phase shifts are invisible in the intensity spectrum, the latter is identical to the one presented in  Fig.~\ref{fig:stepwise}(a) for  constant $\mathbf{B}$ field.
However,  the frequency spectrum differs from Fig.~\ref{fig:stepwise}(b) and shows a large intensity peak at $\omega_0$ and two smaller ones each separated by $2\omega_{QB}$ from the origin. These findings are relevant since fast $\mathbf{B}$-field rotations in $^{57}$Fe-enriched FeBO${}_3$  have been experimentally implemented in Refs.~\cite{Shvydko1996,shvydko:1993,B-rot1995}. However, in these works only the time domain was investigated.

\subsection{Analytical approach \& peak equidistance \label{analysis}}

To obtain a better comprehension of the observed phenomena, we attempt a semi-analytical description of the spectra. The term semi-analytical is used to convey that we do not analytically solve the underlying differential equations, yet we find an analytic function that models the observed time spectra in first order. This function is then analytically Fourier transformed and compared to the actual, discrete transform from the preceding sections.

For a constant magnetic field, the time spectrum dynamics of the scattered field amplitude can be modeled in the first-order approximation by an exponentially decaying sine wave
\begin{equation}
\Omega_{B=\rm{const.}}(L,t) \approx \Omega_0 \cdot \sin(\omega_{QB} t) e^{-\frac{\Gamma}{2}(t)} \Theta(t),
\end{equation}
with a constant amplitude $\Omega_0$. The Heaviside-Theta-function $\Theta(t)$ ensures our spectrum starts at time $t=0$. Note that the first-order approximation assumes that the resulting field corresponds to  single-scattering events. This is applicable here given our choice of small optical thickness $\xi$ corresponding to the thin-sample limit.

\begin{figure}
\centering
\includegraphics[width=0.85\linewidth]{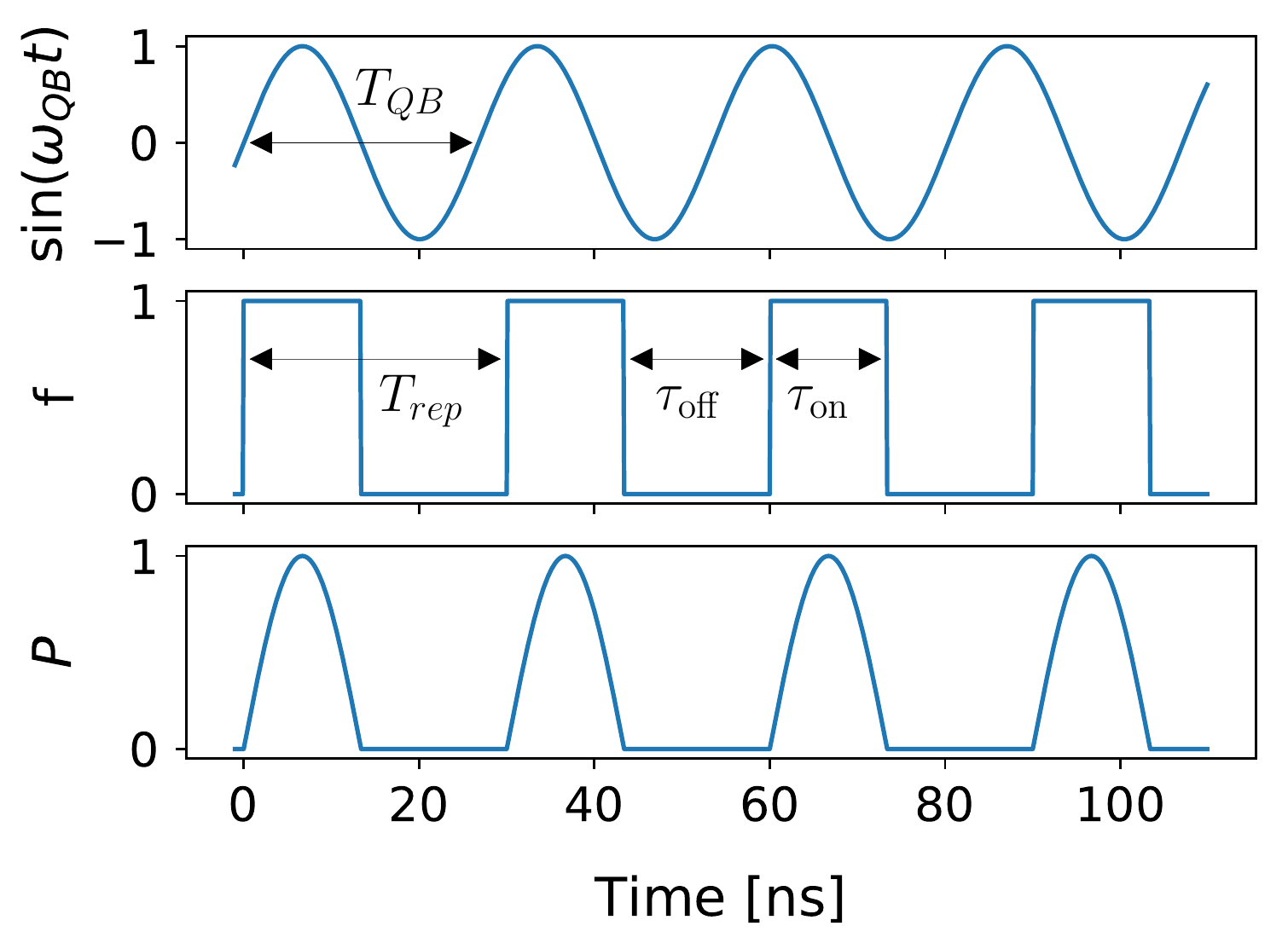}
\caption{Basic components of the analytical model. The combination of the sinusoidal signal $\sin(\omega_{QB}t)$ given via the quantum beat and the periodic sequence $f$ determined by the B-field switching leads to the periodic function $P$.}
\label{fig:analytical-approach}
\end{figure}

The general time spectrum for a non-vanishing storage time $\tau_{\rm{off}}$ and on-time $\tau_{\rm{on}}$ resembles a series of single, cut-off sine-beats
\begin{equation}
P_{single}(t, \tau_{\rm{on}})=\sin(\omega_{QB}t)\, \Pi\left(\frac{t-\tau_{\rm{on}}/2}{\tau_{\rm{on}}}\right),
\end{equation}
where $\Pi$ represents the rectangle function defined in section \ref{B-field}.
The full spectrum can be modeled by sequences of these single beats $P(t,T_{rep},\tau_{\rm{on}}) = \sum_{n=-\infty}^{\infty} P_{single}(t-nT_{rep}, \tau_{\rm{on}})$, e.g.,
\begin{align}
\Omega_{\rm{reg}}(t) &= \Omega_0  \e^{-\frac{\Gamma}{2}(t)} \big[ P(t, 2(\tau_{\rm{on}}+ \tau_{\rm{off}}), \tau_{\rm{on}}) \nonumber \\
&\qquad - P(t-(\tau_{\rm{on}}+ \tau_{\rm{off}}), 2(\tau_{\rm{on}}+ \tau_{\rm{off}}), \tau_{\rm{on}}) \big] \Theta(t), \nonumber \\
\Omega_{\rm{inv}}(t) &= \Omega_0 \e^{-\frac{\Gamma}{2}(t)} P(t, \tau_{\rm{on}}+ \tau_{\rm{off}}, \tau_{\rm{on}})  \Theta(t),
\label{eq:anaTime}
\end{align}
where $P$ is a periodic function in time with period $T_{rep}$. The construction of $P$ starting from its basic components, the quantum beat $\sin(\omega_{QB}t)$ and the periodic sequence $f(t)$ [for this example from Eq.~(\ref{eq:f_regular})], is illustrated in Fig.~\ref{fig:analytical-approach}. Note that since we are only interested in the frequency dependence of the intensity, we neglect a time shift of $(t_0 + \tau_{\rm{off}})$ here in comparison to Eqs.~(\ref{eq:f_regular}) and (\ref{eq:f_inv}). This shift corresponds to an overall phase factor of the electric field that is not relevant for our purposes and could be easily subsequently incorporated into the frequency spectra.
By expanding $P$ as a Fourier series, it is possible to obtain an analytical expression for the Fourier transforms $\tilde{\Omega}_{\rm{reg}}(\omega)$ and $\tilde{\Omega}_{\rm{inv}}(\omega)$,
\begin{align}
\tilde{\Omega}_{\rm{reg}}(\omega) &= \Omega_0 \sum_{n=odd} \frac{2 c_n(2(\tau_{\rm{on}}+\tau_{\rm{off}}),\tau_{\rm{on}})}{i\left(\omega-n\frac{\pi}{\tau_{\rm{on}}+\tau_{\rm{off}}}\right)+\Gamma/2}, \nonumber \\
\tilde{\Omega}_{\rm{inv}}(\omega) &= \Omega_0 \sum_{n=-\infty}^{\infty} \frac{c_n(\tau_{\rm{on}}+\tau_{\rm{off}},\tau_{\rm{on}})}{i\left(\omega-n\frac{2\pi}{\tau_{\rm{on}}+\tau_{\rm{off}}}\right)+\Gamma/2} .
\label{eq:anaFreq}
\end{align}
With the constraint $\tau_{\rm{on}}=m T_{QB}/2$ ($m=1,2,3,\dots$) and the definition $k = T_{rep}/ T_{QB}$ ($T_{rep}\ge\tau_{\rm{on}}$), the Fourier coefficients $c_n$ are given by
\begin{align}
    c_n(T_{rep},\tau_{\rm{on}}) &= \frac{1}{T_{rep}} \int_0^{\tau_{\rm{on}}} P_{single}(t,\tau_{\rm{on}}) \e^{-i n \frac{2\pi}{T_{rep}}t} dt \nonumber \\
    &= \frac{k}{2 (k-n)(k+n)\pi} \left[ 1 - \e^{-i m \pi \left(\frac{n}{k} +1 \right)} \right].
\end{align}

\begin{figure}
\centering
\includegraphics[width=0.85\linewidth]{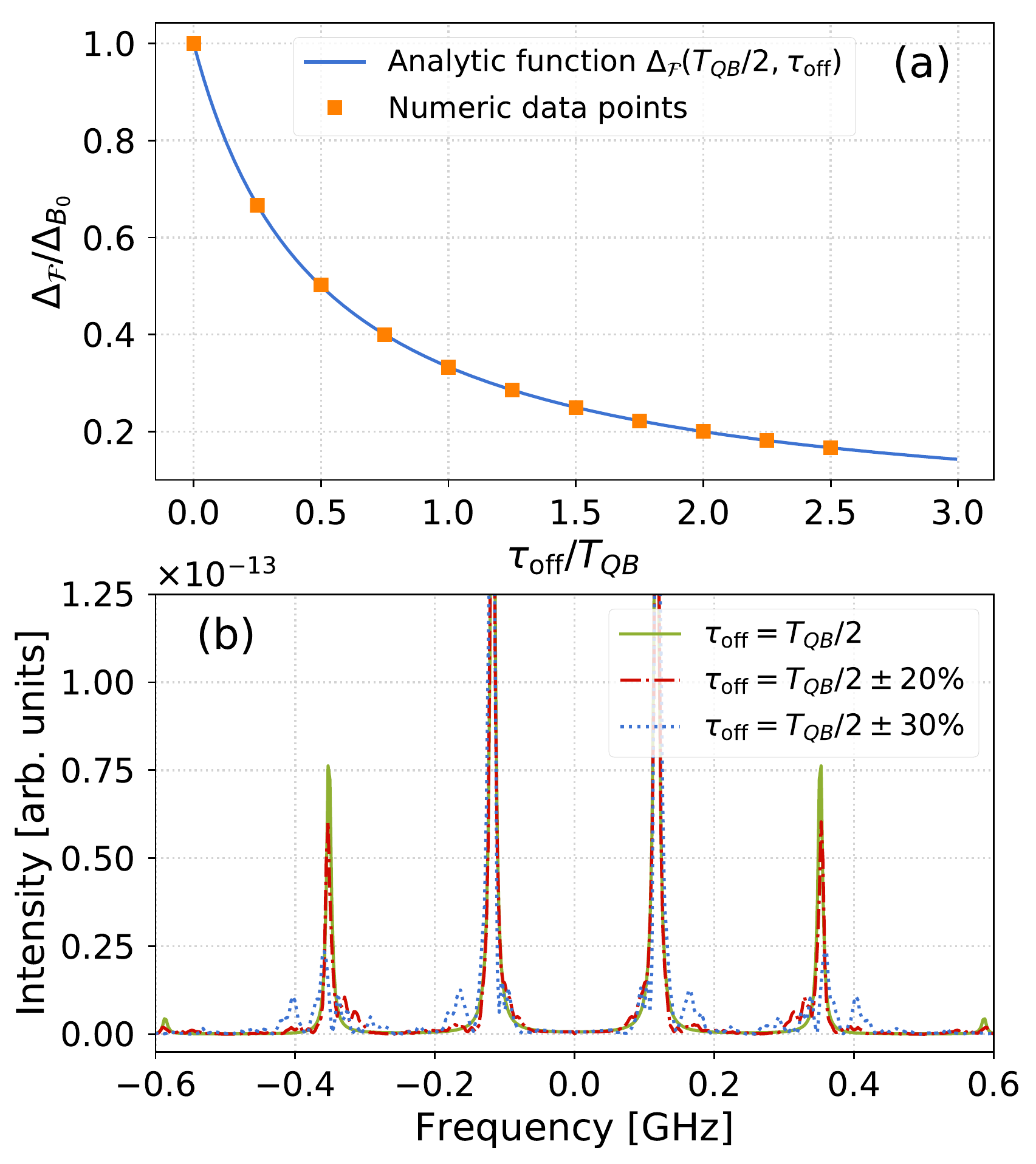}
\caption{(a) Peak distance $\Delta_{\mathcal{F}}$ as a function of the storage time $\tau_{\rm{off}}$. Numeric data points are shown together with the analytic solution given by Eq.~(\ref{eq:delta_F}). (b) Frequency spectra for regular stepwise switching schemes with random variations in $\tau_{\rm{off}}$. For both graphs $\tau_{\rm{on}}=T_{QB}/2$ and $\xi=0.5$ have been considered.}
\label{fig:peak_distance}
\end{figure}

The analytic time spectrum model is in good agreement with the numerical solution from the MBE for small $\xi$. According to Eqs.~(\ref{eq:anaFreq}), the time spectra are given as a summation of individual Lorentzian peaks at frequencies $\omega=(2n+1) \pi/(\tau_{\rm{on}}+\tau_{\rm{off}})$ and $\omega=n 2\pi/(\tau_{\rm{on}}+\tau_{\rm{off}})$ with $n=0,\pm1,\pm2,\dots$ for the regular and inverted stepwise switching schemes, respectively.
For $\tau_{\rm{on}}=T_{QB}/2$ this behavior is clearly reflected in Figs.~\ref{fig:stepwise}--\ref{fig:tau_off_stepwise_inverted}. For both switching schemes, the peak distance evaluates to
\begin{equation}
    \Delta_{\mathcal{F}}(\tau_{\rm{on}},\tau_{\rm{off}}) = \frac{2 \pi}{\tau_{\rm{on}} + \tau_{\rm{off}}} = \frac{1}{2} \Delta_{B_0} \left( \frac{T_{QB}}{\tau_{\rm{on}} + \tau_{\rm{off}}} \right),
    \label{eq:delta_F}
\end{equation}
which is confirmed by numerical results for $\tau_{\rm{on}}=T_{QB}/2$ as shown in  Fig.~\ref{fig:peak_distance}(a). The standard deviation of the presented data points $\Delta_{\mathcal{F}}/\Delta_{B_0}$ lies between 0.09\% and 0.13\% which is a first proof for the equidistance of the occurring  peaks in the frequency domain.

To further check the equidistance we also evaluated the repetition period of the generated sinusoidal pulse train in the time domain, since this periodic pattern is the origin of the comb-like structure in the frequency spectra.
The repetition period has a maximum standard deviation of 0.15 ns among all the considered scenarios with a regular stepwise switching scheme.
We attribute this deviation to the limited accuracy of the numeric solution of the MBE and the contribution beyond first order scattering. To further analyze the effect of limited periodicity, we mimic a non-periodic switching scheme by randomly choosing the storage time $\tau_{\rm{off}}$ within a given interval. In  Fig.~\ref{fig:peak_distance}(b), frequency spectra for a regular stepwise switching scheme are presented where $\tau_{\rm{off}}$ has been randomly drawn from a uniform distribution in the range $T_{QB}/2 \pm 20\%$ and $T_{QB}/2 \pm 30\%$. Comparing the outcomes with $\tau_{\rm{off}}=T_{QB}/2$, one can see that the non-periodic time pattern destroys the comb-like structure in the frequency domain.
While the peaks are still visible for the case $T_{QB}/2 \pm 20\%$, each peak becomes more and more asymmetric and noisy the higher the variation in $\tau_{\rm{off}}$.

\subsection{Experimental feasibility}
The analytic form of the magnetic field modulation implemented in the MBE so far assumes instantaneous switching off and on of a high-intensity magnetic field. This assumption facilitates the calculations and is useful for a first understanding of the time and frequency spectra, but is not realistic from the experimental point of view. The $B=34.4$~T value used for our calculations is the typical value of the magnetic field at the nuclear site in $^{57}$Fe-enriched FeBO${}_3$ crystals. A weak ferromagnet, FeBO${}_3$ is known to allow fast rotation of its magnetization by a weak external magnetic field which aligns the electron spins \cite{Shvydko1996,shvydko:1993,B-rot1995}. While the orientation of the electron spins and the resulting magnetic field at the nucleus can be rotated in less than 4~ns \cite{Shvydko1994.EPL}, its absolute value however cannot be influenced externally.

A different route to achieve the desired magnetic field modulation is to use strong external magnetic fields. The experimental challenges for the control on ns time scale of strong magnetic fields have been previously addressed in Refs.~\cite{Liao:2012,Kong:2014}. The most promising solution involves a material  with no intrinsic nuclear hyperfine splitting like stainless steel Fe$_{55}$Cr$_{25}$Ni$_{20}$. The  challenge is to turn off and on the external magnetic fields of few Tesla on the ns time scale. According to the calculations presented in Ref.~\cite{Liao:2012}, the raising time of the {\bf B} field should be shorter than 50 ns to still observe the storage feature. This could be achieved by using small single- or few-turn coils and a moderate pulse current of approx. 15 kA from low-inductive high-voltage ``snapper'' capacitors \cite{Miura2003}. Another mechanical solution, e.g., the lighthouse setup \cite{Roehlsberger2000} could be used to move the excited target out of and into a region with confined static {\bf B} field. The nuclear lighthouse setup  is based on a rotating sample.  This changes the direction of the coherently emitted photon which is always in the forward direction with respect to the
sample, thus explaining the name ``lighthouse effect''. The rotation can be used to bring the sample in and outside a region with confined static magnetic field. The switching time is then given by the time needed for the rotation of the sample from the edge of the confined magnetic field region to the outside, magnetic-field free region.
A rotor with rotational frequencies $R$ of up to 70 kHz and a diameter of few mm \cite{Roehlsberger2000} is fast enough to rotate the sample
out of a depth of few $\mu$m in a few tens of ns.

\begin{figure}[t!]
\includegraphics[width=\linewidth]{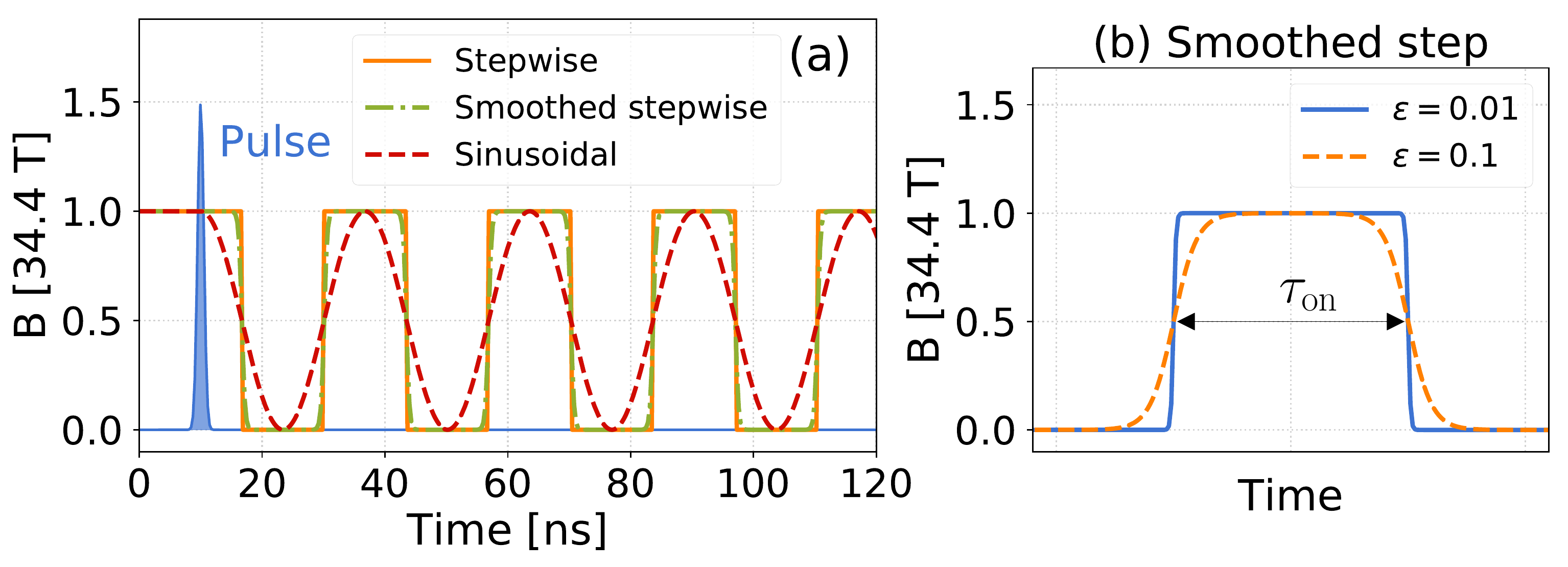}
\caption{(a) Magnetic field modulations: stepwise (orange solid line), smoothed stepwise (green dash-dotted line) with $\varepsilon=0.5$ and sinusoidal (red dashed line). (b) Examples of smoothed stepwise functions as in Eq.~(\ref{smoothfct}) with $\varepsilon=0.01$ and $\varepsilon=0.1$. }
\label{fig:switchings}
\end{figure}

For both scenarios, the switching between $B_0$ and 0 is not a step-wise one. We therefore investigate the effect of more smoothed magnetic switchings with a finite, experimentally more feasible slope  as well as a sinusoidal $\mathbf{B}$-field modulation, as shown in Fig. \ref{fig:switchings}. These sequences allow for a slower $\mathbf{B}$-field fall and rise for instance in the form of a hyperbolic function smoothed step function (see also Ref.~\cite{Rostovtsev2010}) according to
\begin{equation}
    \begin{split}
    \Pi_{smooth}(x) = 
 \frac{1}{2} \left[1+\tanh\left(\frac{x+1/2}{\varepsilon}\right) \right]\Theta(-x) \\ 
+  \frac{1}{2} \left[1-\tanh\left(\frac{x-1/2}{\varepsilon}\right) \right]\Theta(x)
    \end{split}
\label{smoothfct}
\end{equation}
with smoothing radius $\varepsilon$.
Considering this definition a smoothed version of the regular stepwise switching sequence can be obtained by substituting $\Pi_{smooth}$ for $\Pi$ in Eq.~(\ref{eq:f_regular}).
Apart from a hyperbolic function smoothed step, the magnetic field can be considered to oscillate sinusoidally with the quantum beat frequency,
\begin{equation}
    f(t) = \Theta(\tilde{t}_0-t) + 1 - \frac{1}{2}\, \Theta(t-\tilde{t}_0) \sin[ \omega_{QB} (t - t_0) ],
\end{equation}
where $\tilde{t}_0 = t_0 - T_{QB}/4$.
In this case, one degree of freedom is lost by introducing the restriction  $\tau_{\rm{off}}=\tau_{\rm{on}}$.

\begin{figure}
\centering
\includegraphics[width=0.85\linewidth]{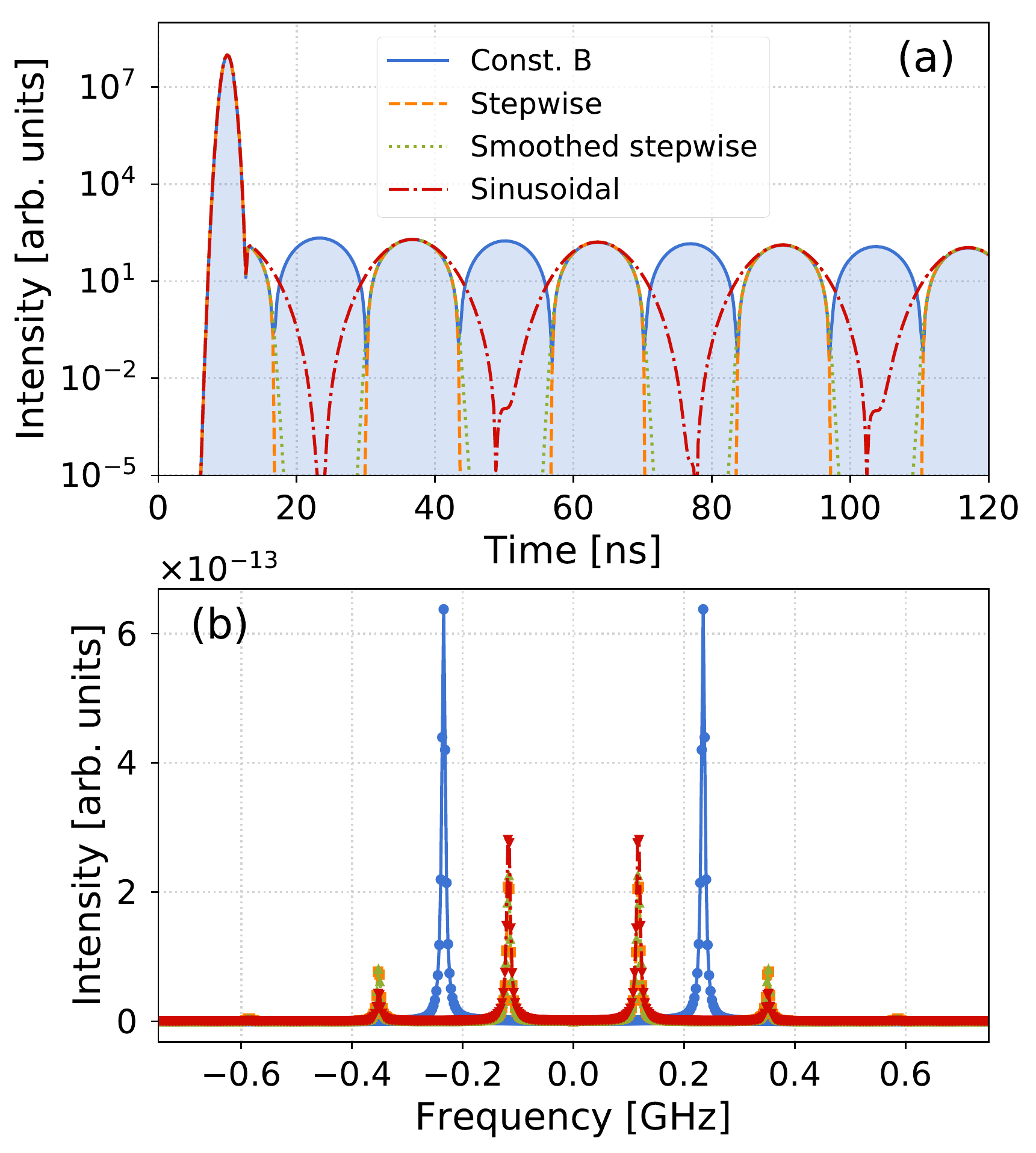}
\caption{(a) Time and corresponding frequency spectra (b) for abrupt (orange dashed line), smoothed (green dotted line) and sinusoidal (red dashed-dotted line) magnetic switching sequences in comparison to the case of a constant magnetic field (blue solid line). The smoothing radius for the hyperbolic function smoothed step  was set to $\varepsilon=0.5$. For all cases we considered $\tau_{\rm{on}}=\tau_{\rm{off}}=T_{QB}/2$ and $\xi=0.5$.}
\label{fig:comp_switching_schemes}
\end{figure}

In Fig.~\ref{fig:comp_switching_schemes}, time and frequency spectra for stepwise, smoothed stepwise and sinusoidal switching schemes are shown in comparison to results for a constant magnetic field. For all presented cases we considered $\tau_{\rm{on}}=\tau_{\rm{off}}=T_{QB}/2$ and $\xi=0.5$.
Except for slight intensity deviations, the main characteristics of the frequency combs are maintained since the periodicity in the time domain is sustained.
Even the sinusoidal signal, which is likely to be the simplest to experimentally implement, shows no significant frequency spectrum distortion. This supports the experimental realization of magnetic field modulation with generation of x-ray frequency-comb-like structure in the near future.

\section{Discussion}
Frequency and phase modulation of x-ray photons has been so far typically realized via mechanical vibrations or displacements of the sample. Here, we have introduced a mechanical-free solution which relies on the periodic switching of an external magnetic field to produce a comb-like  frequency  spectrum of the resonantly scattered x-ray photons. Typically calculated spacings for the peaks are on the level of 0.1~GHz, corresponding to switching periods commensurate with the hyperfine splitting of the nuclear transition and in turn with the intensity of the applied magnetic field. The peak widths are slightly larger than the natural linewidth of 7~MHz, due to weak collective effects for the chosen optical thickness $\xi=0.5$. The key ingredient for the magnetic field switching is the periodicity of the modulation and not its exact shape, which facilitates the experimental implementation of this scheme.

Analogous to optical frequency combs, we anticipate that the proposed x-ray frequency modulation may be of interest in the fields of x-ray frequency measurements or quantum information processing. Typical frequency spectrum recording is achieved by employing a single-line absorber on a M\"ossbauer drive behind the nuclear target. The frequency resolution is approximatively given by the natural linewidth of the absorber, in the case of  $^{57}$Fe~1/141 ns=7 MHz. With the spacing of the generated frequency comb externally controlled by the intensity of the magnetic field, a setup can be designed to determine the resonant frequency of a given system by successive fluorescence detection for several $\mathbf{B}$-field values. Allegedly, the draw-back of such a scenario is the rather inflexible range over which the x-ray frequency comb can be generated, limiting the cases for which such measurements can be performed.

The potential of optical frequency combs for generation of highly multi-mode quantum states for quantum information processing and measurement-based
quantum computing has been repeatedly highlighted in recent years \cite{Menicucci2008,Roslund2014,Gerke2015}. Down-conversion of frequency combs is used to generate highly entangled quantum networks, also known as cluster states \cite{Menicucci2006}, that  can be used for quantum information processing. The key ingredient in this scenario is bipartite entanglement which is generated in the process of down-conversion. For the x-ray comb-like frequency  modulation generated via magnetic switching, SR pulses can typically only deliver a single photon which entangles the field modes with different energies available within the comb. While such single-photon entanglement \cite{LeeKim,Enk2005} has been discussed in the context of quantum information processing, its uses are more limited compared to bipartite entanglement. For several x-ray  photons distributed over the comb energies, or ideally at least one photon within each spectral peak, stronger  resonant driving is required.
 Using a NFS setup driven by XFEL pulses with much stronger brilliance than SR \cite{Chumakov2017} such a scenario is conceivable.

\section*{Acknowledgements}
XK and AP are  part of and were supported by the DFG
Collaborative Research Center SFB 1225 (ISOQUANT).

%


\bibliography{bibliography}{}

\end{document}